\documentclass[%
superscriptaddress,
twocolumn,
showpacs,
amsmath,amssymb,
aps,
floatfix,
showkeys,
]{revtex4-1}

\usepackage[dvips]{graphicx}
\usepackage{dcolumn}
\usepackage{bm}
\usepackage{amsmath}
\usepackage{amssymb}
\usepackage{multirow}
\usepackage{color,colortbl}
\usepackage{moreverb}
\usepackage{booktabs}
\usepackage{tabularx}
\usepackage[english]{babel}
\usepackage[utf8]{inputenc}
\usepackage{physics}
\usepackage[version=4]{mhchem}
\usepackage{enumerate}
\usepackage{footnote}
\usepackage{tablefootnote}

\usepackage[backref=true,colorlinks=true,citecolor=blue,filecolor=black,urlcolor=black]{hyperref}

\usepackage[normalem]{ulem}

\usepackage{xargs}                      
\usepackage[xelatex,dvipsnames]{xcolor}  
\def\bra#1{\mathinner{\langle{#1}|}}
\def\ket#1{\mathinner{|{#1}\rangle}}

\newcommand{\etal}{\textit{et al.\/}}

\newcommand{\eg}{e.\,g.,}
\newcommand{\uB}{\ensuremath{\mu_{\text{B}}}}

\definecolor{Gray}{gray}{0.85}
\definecolor{LightCyan}{rgb}{0.88,1,1}

\begin{document}

\title{Comparative study of electronic and magnetic properties of \emph{M}Pc (\emph{M} = Fe, Co) molecules physisorbed on 2D \ce{MoS2} and graphene}

\author{Soumyajyoti Haldar}
\altaffiliation[Current Address: ]{Institute of Theoretical Physics and Astrophysics, University of Kiel, Leibnizstrasse 15, 24098 Kiel, Germany}
\affiliation {Division of Materials Theory, Department of Physics and Astronomy, Uppsala University, Box-516, SE 75120, Sweden}

\author{Sumanta Bhandary}
\affiliation {Centre de Physique Theorique (CPHT), Ecole Polytechnique, 91128 Palaiseau cedex, France}

\author{Hakkim Vovusha}
\altaffiliation[Current Address: ]{King Abdullah university of science and technology ( KAUST), Physical Science and Engineering Division ( PSE), Thuwal 23955-6900, Saudi Arabia}
\affiliation {Division of Materials Theory, Department of Physics and Astronomy, Uppsala University, Box-516, SE 75120, Sweden}

\author{Biplab Sanyal}
\email[Email: ]{Biplab.Sanyal@physics.uu.se}
\affiliation{Division of Materials Theory, Department of Physics and Astronomy, Uppsala University, Box-516, SE 75120, Sweden}

\date{\today}

\begin{abstract}
In this paper, we have done a comparative study of electronic and magnetic properties of iron phthalocyanine (FePc) and cobalt phthalocyanine (CoPc) molecules physisorbed on monolayer of \ce{MoS2} and graphene by using density functional theory. Various different types of physisorption sites have been considered for both surfaces. Our calculations reveal that the \emph{M}Pc molecules prefer the S-top position on \ce{MoS2}. However, on graphene, FePc molecule prefers the bridge position while CoPc molecule prefers the top position.  The \emph{M}Pc molecules are physisorbed strongly on the \ce{MoS2} surface than the graphene ($\sim$ 2.5 eV higher physisorption energy). Analysis of magnetic properties indicates the presence of strong spin dipole moment opposite to the spin moment and hence a huge reduction of effective spin moment can be observed. Our calculations of magnetic anisotropy energies using both variational approach and $2^{nd}$ order perturbation approach indicate no significant changes after physisorption. In case of FePc, an out-of-plane easy axis and in case of CoPc,  an in-plane easy axis can be seen. Calculations of work function indicate a reduction of \ce{MoS2} work function $\sim$ 1 eV due to  physisorption of \emph{M}Pc molecules while it does not change significantly in case of graphene.    

\end{abstract}

\keywords{Metal phthalocyanine, physisorption on \ce{MoS2}, physisorption on graphene, magnetic properties, work function}
\maketitle


\section{\label{sec:intro}Introduction}

The feasibility of spin and charge degrees of freedom manipulation has made molecular electronics and spintronics an interesting field of research. Metal-organic molecules with a metal atom center is a key prospect in this field. The efficient spin manipulation has made these molecules an attractive choice for various kind of spin dependent electronics applications.~\cite{Leoni2011,Wende2007,Dediu2009,sumantaprb}
Metal phthalocyanines (\emph{M}Pc) are porphyrinoid macrocyclic complexes and symmetric organic molecules with a metal atom (\emph{M}) at the center surrounded by four bonded N atom and four non bonded N atom.
The phthalocyanine molecules do not have any out-of-plane ligands. Hence, they can remain planar on a surface and enhance the surface-molecular interactions. 
These molecules, especially transition metal phthalocyanines have application prospects in various scientific fields {\eg} photovoltaics, organic solar cells, molecular electronics and spintronics etc.~\cite{Martinez2010,Ishikawa2010,Bogani2008,Liu2013,Strozecka2012,Fu2007,ANIE:ANIE201108963,B212621D} Long range structural ordering on metallic surface and influence of surface on the electronic properties has also been experimentally realized.~\cite{Betti2010}

However, for many of these applications, the $M$Pc molecules need to be absorbed or hosted on some kind of molecule surfaces. In this context, the use of stable ultra thin atomic materials are an automatic choice for future devices. Two dimensional materials like graphene are one of the automatic choices in this regard where it is already being used quite significantly.~\cite{netormp,Geim2007b,Geim2009}
Apart from graphene, \ce{MoS2} is also another choice of materials  due to its interesting properties. Two dimensional (2D) \ce{MoS2} being an ultra-thin semiconductor with unique electronic and optical characteristics, has potential application in optoelectronics, in fluorescence imaging, in photo catalysis, in solar cells, in valley electronics, as photodynamic, photothermal materials and even in biomedicine as antibacterial agents.~\cite{p1,p2,p3,p4,p5,p6,p7} Monolayer \ce{MoS2} has a direct band gap of 1.9 eV, which is suitable to absorb visible light.~\cite{Haldar2015,p8} Tuning of electronic and optical properties of \ce{MoS2} by means of molecular charge transfer using organic molecules adsorption have also been reported in recent theoretical studies.~\cite{p9} Effect of \ce{CuPc} and \ce{TiOPc} physisorption in optical properties on 2D \ce{MoS2} have also been investigated recently.~\cite{Choudhury2017}

Magnetic anisotropy energies are also quite important factor related to spintronics applications. For example, to increase storage density one needs to have an out-of-plane magnetization and magnetic molecules can play an important role into it. Strong magnetic anisotropy for metallic atoms in presence of graphene have been reported in recent studies.~\cite{kandpal,porter} Graphene induced magnetic anisotropy in two dimensional phthalocyanine network also has been reported recently.~\cite{Lisi2015}
Motivated by these studies, we have also investigated magnetic anisotropy energies of FePc and CoPc magnetic molecules  physisorbed on \ce{MoS2} and graphene. 

In low dimensional system such as clusters, organometallic molecules with metal center, spin-dipole interaction ($\langle  T_z \rangle$) can become very important.~\cite{hubert,sumantaprl,gambardella}. These spin-dipole contributions, originating from non isotropic spin densities, may have opposite sign from the spin moment and hence reducing the effective moment ($m_{eff}$ = $m_s$ +7$\langle  T_z \rangle$). This effective moment can be measured in X-ray magnetic circular dichroism (XMCD) experiments.~\cite{heikeprb}

In this paper, we have done a comparative study of electronic and magnetic properties of metal phthalocyanine molecules 
physisorbed on monolayer of graphene and 2D \ce{MoS2} sheet by using density functional theory. Iron phthalocyanine (FePc) and cobalt phthalocyanine (CoPc) molecules were chosen as the representatives of metal phthalocyanine molecules.  Our calculations reveal that the \emph{M}Pc molecules physisorbed strongly on the \ce{MoS2} surface than the graphene. The stable physisorption sites are different for \ce{MoS2} and graphene. We have compared the density of states with the free molecular states.  Changes in work function, magnetic anisotropies and spin dipolar contribution due to the physisorption are also discussed in details. The plan of the paper is as follows. In the next section we will present the computational details followed by results and discussion in section~\ref{sec:results}. In result and discussion section, we will first discuss about the structural properties [\ref{subsec:structure}] and then it will be followed by discussion of electronic structure [\ref{subsec:elstructure}], work function [\ref{subsec:workfunc}], spin dipole contributions [\ref{subsec:spindipole}] and magnetic anisotropy [\ref{subsec:mae}]. The conclusions are discussed in section~\ref{sec:conclusion}

\section{\label{sec:method}Computational Details}
All the calculations have been performed in a monolayer supercell of \ce{MoS2} and graphene.  The supercells of \ce{MoS2} and graphene are generated by repeating the primitive cells respectively by 9 and 12  times in both $a$ and $b$ directions. The \emph{M}Pc ($M$=Fe, Co) molecules was placed on top of these supercells. It is extremely important to reduce the interaction between  periodic images of the physisorbed molecules and hence such a large size of supercell is essential. A vacuum of 20 {\AA} is included to avoid the effect of vertical interaction.  All the calculations have been performed using plane-wave based density functional code \textsc{vasp}. \cite{vasp}  The generalized gradient approximation of Perdew, Burke and Ernzerhof ~\cite{PBE,PBEerr} has been used  for the exchange-correlation potential.  All the structures have been fully relaxed using the conjugate gradient method with the forces calculated using the Hellman-Feynman theorem. We have used the energy and the Hellman-Feynman force thresholds at 10$^{-4}$ eV and 10$^{-2}$ eV/{\AA} respectively. We have used 500 eV cutoff energy to truncate the plane waves. A 3 $\times$ 3 $\times$ 1 Monkhorst-Pack $k$-grid was used for all our calculations. To address the problem of electron correlation effect in the narrow $d$ states of metal atoms (Fe, Co), we have used a GGA+U approach following the formalism proposed by Dudarev \etal~\cite{Dudarev} A Coulomb interaction term is added in according to the mean field Hubbard U formalism.~\cite{Dudarev,uj1,uj2} The value of the exchange parameter, J, was chosen to be 1.0 eV. The value of Coulomb parameter, U, was chosen to be 4.0 eV and 6.0 eV, respectively, for FePc and CoPc.~\cite{Wehling:2011dw,Eelbo:2013ft} These values were chosen as 
they correctly reproduce the electronic structure and magnetic properties of FePc and CoPc in gas phase. 

The van der Waals interactions between the \emph{M}Pc molecules and the \ce{MoS2} or graphene layer originate due to the fluctuating charge distribution and hence cannot be described by PBE functionals alone. Hence for all the calculations, we have included the van der Waals effect by adding a correction to the conventional Kohn-Sham DFT energy through pair wise force fields following the method of Tkatchenko-Scheffler.~\cite{vdw-ts} 

The physisorption energy $E_a$ for metal phthalocyanine molecule physisorbed on \ce{MoS2} or graphene is defined as follows, 
\begin{equation}
\label{eq:physisorption}
 E_{a} = \big[E_{\rm \ce{layer}} + E_{\rm \emph{M}Pc}\big]-E_{\rm layer + \emph{M}Pc}
\end{equation}
where, 
\begin{enumerate}[i]
	\item $E_{\rm layer}$ is the total energy of either \ce{MoS2} or graphene supercell
	\item $E_{M\rm Pc}$ is the total energy of the \emph{M}Pc molecules in the gas phase
	\item $E_{\rm layer + \emph{M}Pc}$ is the total energy of \emph{M}Pc physisorbed on \ce{MoS2} or graphene supercell system.
\end{enumerate}

\section{\label{sec:results}Results and discussions}
\subsection{\label{subsec:structure}Structural properties}
We will begin our discussion of results from the analysis of structural properties and the energetics of the systems studied. We will first discuss the results of \emph{M}Pc physisorption on \ce{MoS2} followed by the results of physisorption on graphene. 

\begin{table}[htbp]
	\caption{Physisorption energy ($E_a$), total magnetic moment (\uB) and distance of $M$ atom (D$_h$) from \ce{MoS2} for $M$Pc physisorbed on MoS$_2$. \\}
	\label{tab:energy_mag_mos2}
	\begin{tabular}{c | c c c}
		\rowcolor{Gray}
		FePc@ & E$_a$& \uB & D$_h$ \\
		\rowcolor{Gray}
		\ce{MoS2} &  (eV) & (total) & ({\AA}) \\ \hline
		\rowcolor{LightCyan}
		\textbf{S-Top} & \textbf{6.21} & \textbf{2.0} & \textbf{3.18} \\
		Bridge 	& 6.06 & 2.0 & 3.42 \\
		Hex & 6.03 & 2.0 & 3.37 \\
		Mo-Top & 5.98 & 2.0 & 3.39 \\ \hline
	\end{tabular}
	\begin{tabular}{c | c c c}
				\rowcolor{Gray}
		CoPc@ & E$_a$& \uB & D$_h$ \\
				\rowcolor{Gray}
		\ce{MoS2} &  (eV) & (total) & ({\AA}) \\ \hline
		\rowcolor{LightCyan}
		\textbf{S-Top} & \textbf{6.18} & \textbf{1.0} & \textbf{3.21} \\
		Bridge 	& 6.05 & 1.0 & 3.42 \\
		Hex & 6.01 & 1.0 & 3.38 \\
		Mo-Top & 5.96 & 1.0 & 3.40 \\ \hline
	\end{tabular}
\end{table}

To find out the most energetically favorable  physisorbed position, we have considered different possible positions of the \emph{M}Pc molecules. These different positions of the molecule can be distinguished by the position of metal atom center with respect to the 
underlying surface layer of \ce{MoS2} or graphene. For \ce{MoS2}, there are four 
possible positions, which are i) S-Top -- when the metal atom center is at the 
top of S atom, ii) Mo-Top -- when the metal atom center is at the top of Mo atom 
iii) Bridge -- when the metal atom center is at the bridge position of two top layer 
S atom of \ce{MoS2} and iv) Hex  -- when the metal atom center is at the hexagonal 
position formed by \ce{MoS2}. In the case of graphene, three positions are 
possible. Following the above terminology, these positions are i) 
C-Top, ii) Bridge and iii) Hex. In our calculations, \emph{M}Pc molecules are 
placed on the surfaces with a particular $xy$ position and we have not 
considered any rotation of the molecules in the $xy$ plane.  

\begin{figure}[tbp]
\begin{center}
\includegraphics[scale=0.6]{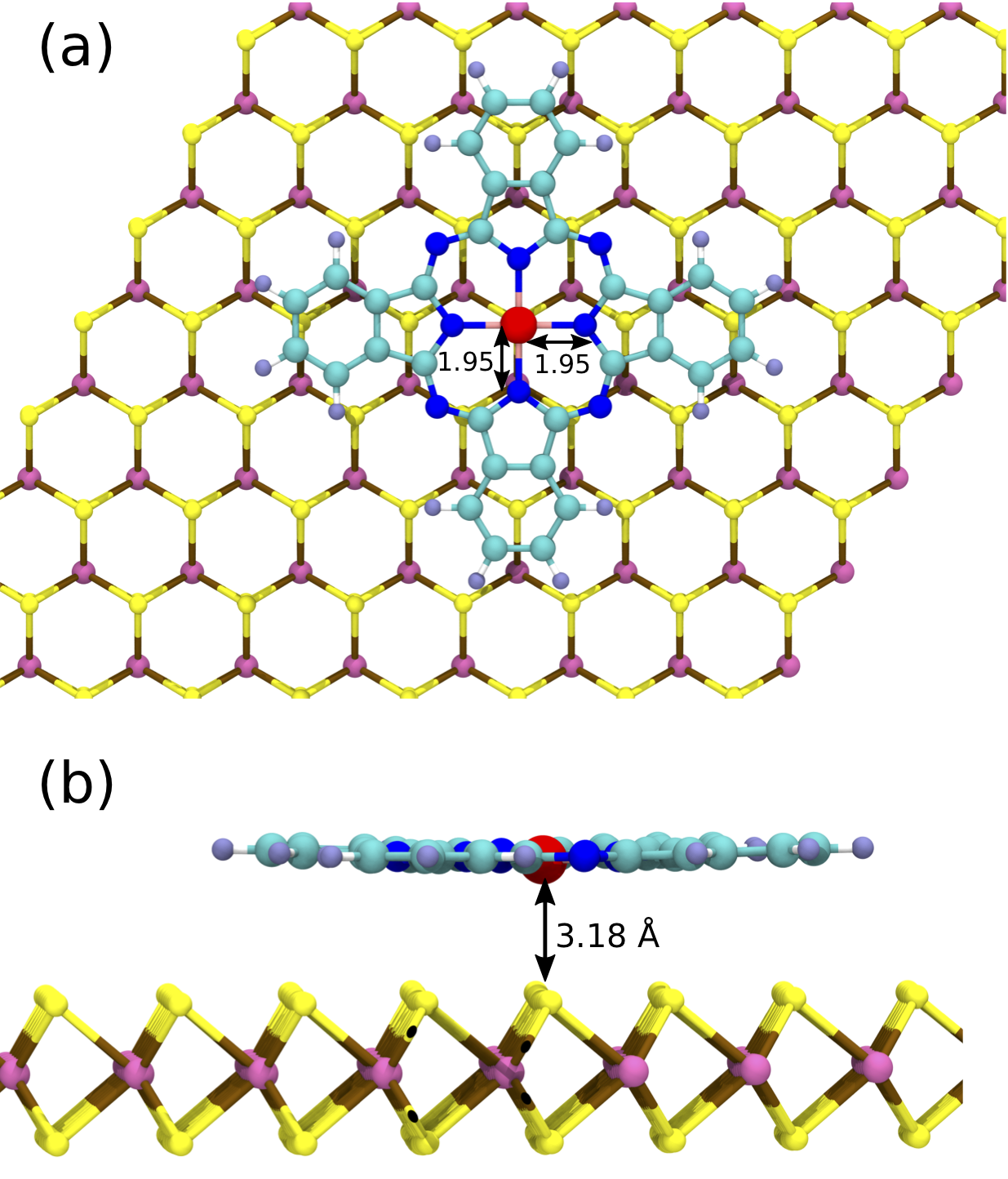}
\caption{(Color online) Closeup view of optimized geometry of FePc molecule on MoS$_2$. (a) Top view and (b) side view. Numbers in (a) and (b) show the Fe-N bond lengths and vertical distance between the metal center of the molecule and the \ce{MoS2} layer respectively. }
\label{fig:fepc-geom}
\end{center}
\end{figure}
\begin{figure}[tbp]
\begin{center}
\includegraphics[scale=0.6]{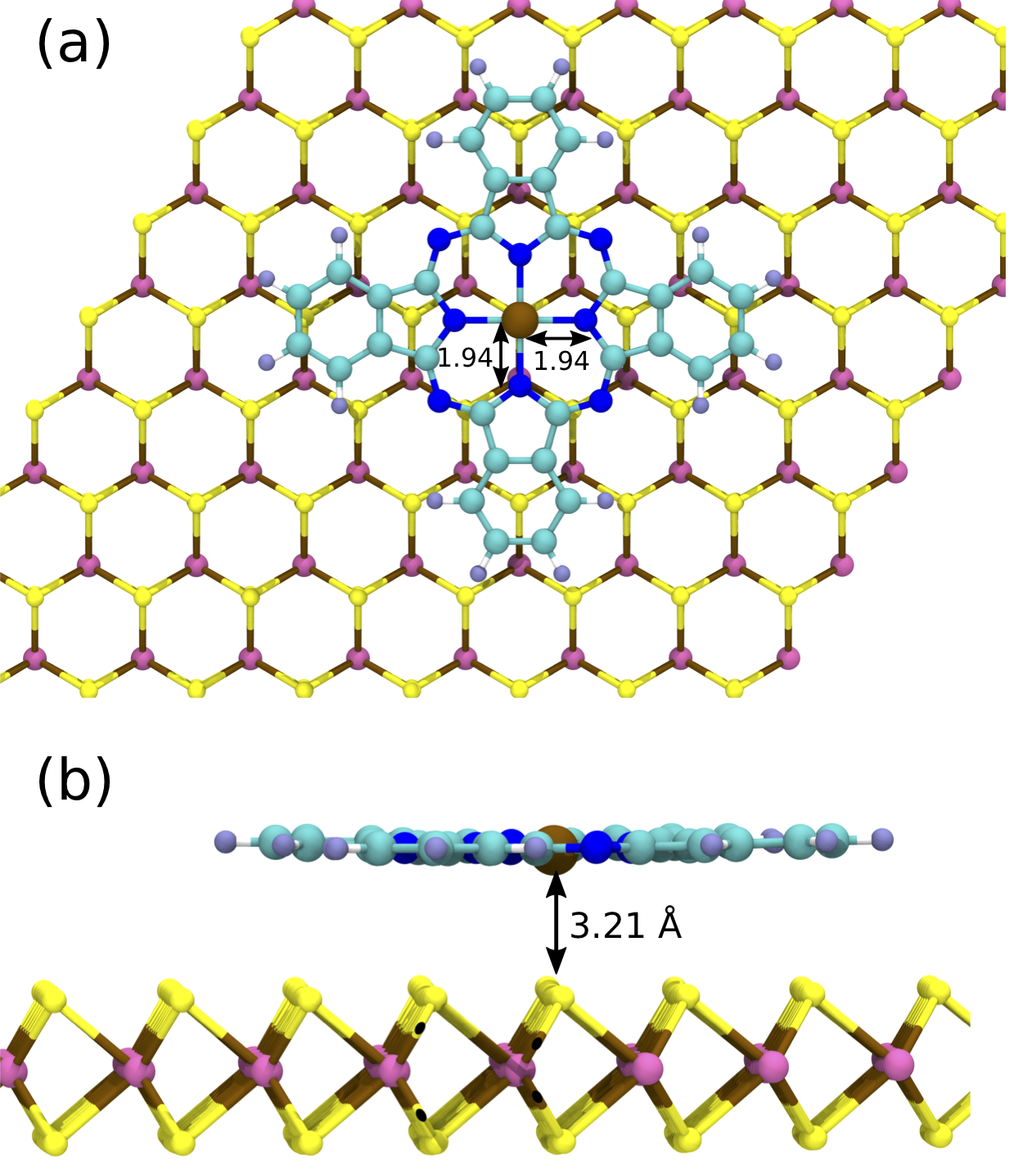}
\caption{(Color online) Closeup view of optimized geometry of CoPc molecule on MoS$_2$. (a) Top view and (b) side view. Numbers in (a) and (b) show the Co-N bond lengths and vertical distance between the metal center of the molecule and the \ce{MoS2} layer respectively.}
\label{fig:copc-geom}
\end{center}
\end{figure}
In table~\ref{tab:energy_mag_mos2}, we have shown the physisorption energies ($E_a$), total magnetic moment ($\mu_B$) and vertical distance between the metal center and the \ce{Mos2} layer for all of the four FePc and CoPc physisorbed structures described before. From the analysis of physisorption energy (see table.~\ref{tab:energy_mag_mos2}), it is clear that for both FePc and CoPc, the energetically most stable structure is when the metal center of the \emph{M}Pc molecules reside exactly at the top of the S atom. Fig.~\ref{fig:fepc-geom} and Fig.~\ref{fig:copc-geom} show the close up of  schematic figure of energetically most stable structures of FePc and CoPc physisorption on \ce{MoS2} respectively. For both cases, all four metal atoms to nitrogen distance remain identical -- 1.95 {\AA} for FePc and 1.94 {\AA} for CoPc maintaining a $D_{4h}$ symmetry. However, the vertical height from the \ce{MoS2} layer are different for the two system. The vertical distance between FePc and CoPc molecule with \ce{MoS2} layer is 3.18 {\AA} and 3.21 {\AA} respectively. 

We have also calculated the $M$Pc physisorption on graphene to compare the same with the $M$Pc physisorption on \ce{MoS2}. To find out the most energetically favorable physisorption sites, we have optimized three different structures of $M$Pc physisorbed on graphene. As stated previously, these positions are different according to the position of the metal atoms of the $M$Pc relative to the C atom from the graphene layer.  
\begin{table}[htbp]
	\caption{Physisorption energy ($E_a$), total magnetic moment (\uB) and distance of $M$ atom (D$_h$) from graphene for $M$Pc physisorbed on graphene. \\}
	\label{tab:energy_mag_gr}
	\begin{tabular}{c|c c c}
				\rowcolor{Gray}
		FePc@ & E$_a$& \uB & D$_h$ \\
				\rowcolor{Gray}
		\ce{Gr} &  (eV) & (total) & ({\AA}) \\ \hline
		Top & 3.38 & 2.0 & 3.37 \\
				\rowcolor{LightCyan}
		\textbf{Bridge} 	& \textbf{3.40} & \textbf{2.0} & \textbf{3.34} \\
		Hex & 3.36 & 2.0 & 3.43 \\ 
		\hline
	\end{tabular}
	\begin{tabular}{c|c c c}
				\rowcolor{Gray}
		CoPc@ & E$_a$& \uB & D$_h$ \\
				\rowcolor{Gray}
		\ce{Gr} &  (eV) & (total) & ({\AA}) \\ \hline
				\rowcolor{LightCyan}
		\textbf{Top} & \textbf{3.39} & \textbf{1.0} & \textbf{3.39} \\
		Bridge	& 3.38 & 1.0 & 3.34 \\
		Hex & 3.35 & 1.0 & 3.43 \\ \hline
	\end{tabular}
\end{table}
The calculated physisorption energy ($E_a$), total magnetic moment ($\mu_B$) and vertical distance between the metal center and graphene layer (D$_h$) are tabulated in Table~\ref{tab:energy_mag_gr}. The analysis of physisorption energies shows that for graphene, the most favorable site for FePc molecule physisorption is when the Fe metal center resides at the bridge position (figure not shown). This is quite different from the physisorption of iron porphyrin (FeP) molecule on graphene, where the energetically favorable physisorption position is on the top of C atom.~\cite{sumantansr}  However, for CoPc molecule the most stable physisorption site on graphene is the top site, where the metal atom is on top of the carbon atom. A comparison of results listed in table~\ref{tab:energy_mag_mos2} and table~\ref{tab:energy_mag_gr} show an interesting contrast between the two cases of \emph{M}Pc molecules physisorption on \ce{MoS2} and graphene. It can be clearly observed that the physisorption energies of \emph{M}Pc physisorption are quite high on \ce{MoS2} as compare to graphene. The high values of physisorption energy in \ce{MoS2} are due to stronger hybridization of \emph{M}-$d$ states with the S-$p_z$ orbitals of \ce{MoS2}. On graphene, the hybridization is relatively weak and hence the physisorption energies are much lower. The different strengths of hybridization are also reflected on the vertical distances of \emph{M}Pc molecules from the \ce{MoS2} and graphene surface. For energetically most favorable physisorption position, the vertical distance between molecule metal center and graphene is $\sim$0.20 {\AA} higher as compared to the vertical distance between molecule metal center and \ce{MoS2}. 

It is also interesting to note that the physisorption energy differences between different position of \emph{M}Pc molecules on graphene are quite small ($\sim$ 0.02 eV). However, on \ce{MoS2} the physisorption energy differences between the most stable physisorption site and next probable site are substantially higher as compared to graphene ($\sim$ 0.14 eV). The reason of this behavior lies in the fact that graphene is planar where as \ce{MoS2} has S and Mo atoms in different layers.  


Our analysis of the structural properties and energetics suggests that \emph{M}Pc molecules will be quite mobile on graphene surface, but they will be strongly physisorbed on \ce{MoS2} surface. Thus, it will be possible to use 2D surface of \ce{MoS2} to isolate single \emph{M}Pc molecules. 

\subsection{\label{subsec:elstructure}Electronic structure}
\begin{figure}[tbp]
\begin{center}
\includegraphics[scale=0.4]{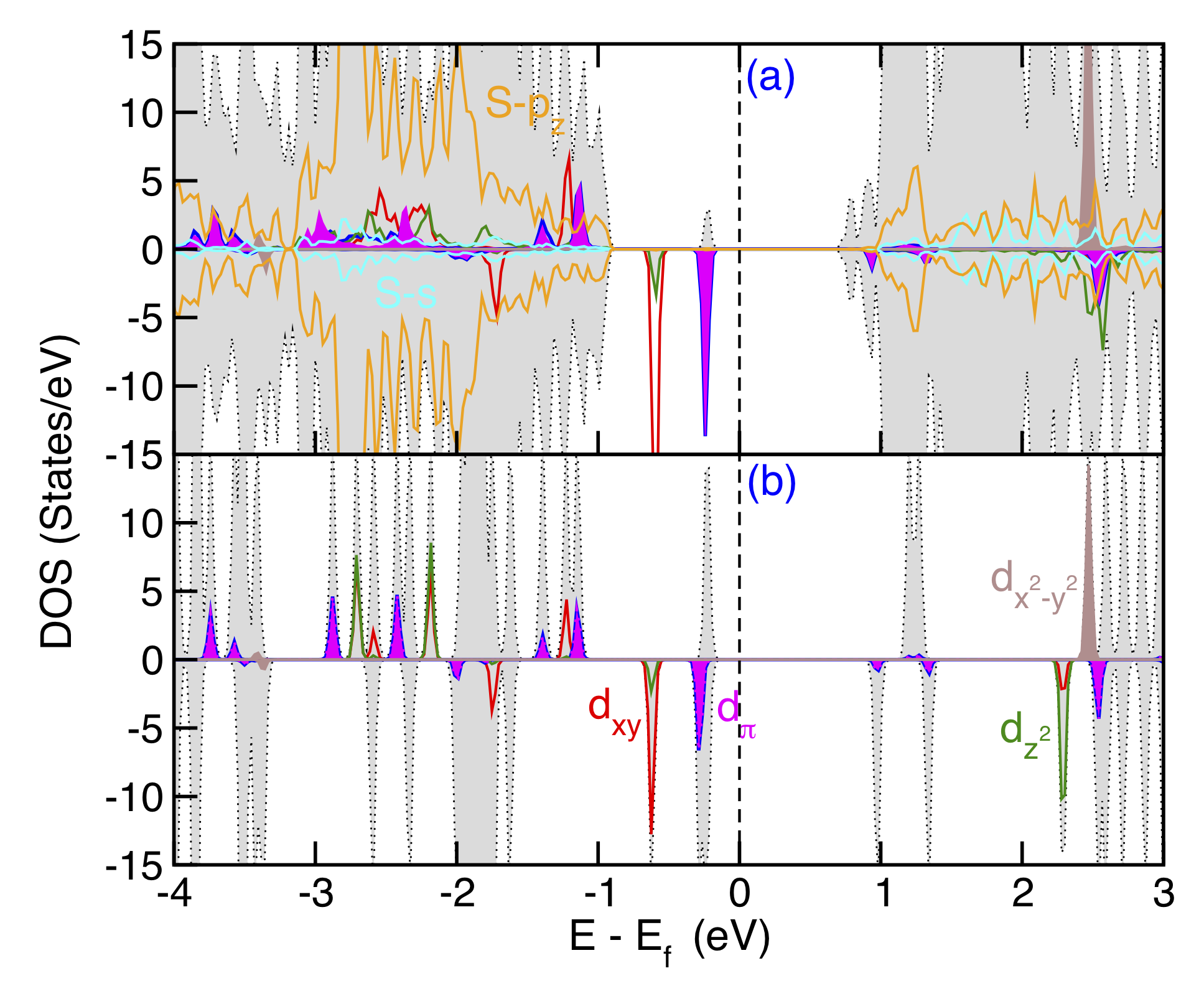}
\caption{(Color online) (a) Total density of states  (grey shaded region) of FePc physisorbed  on MoS$_2$, $m_l$ projected Fe-$d$ states of FePc and $s$, $p_z$ orbitals of top layer S atoms of \ce{MoS2}. The total density of states is scaled down by a factor of 10. (b)  Total density of states  (grey shaded region) and $m_l$ projected Fe-$d$ states for FePc in gas phase.\\~~\\  }
\label{fig:fepc-dos}
\end{center}

\end{figure}
\begin{figure}[tbp]
\centering
\includegraphics[scale=0.4]{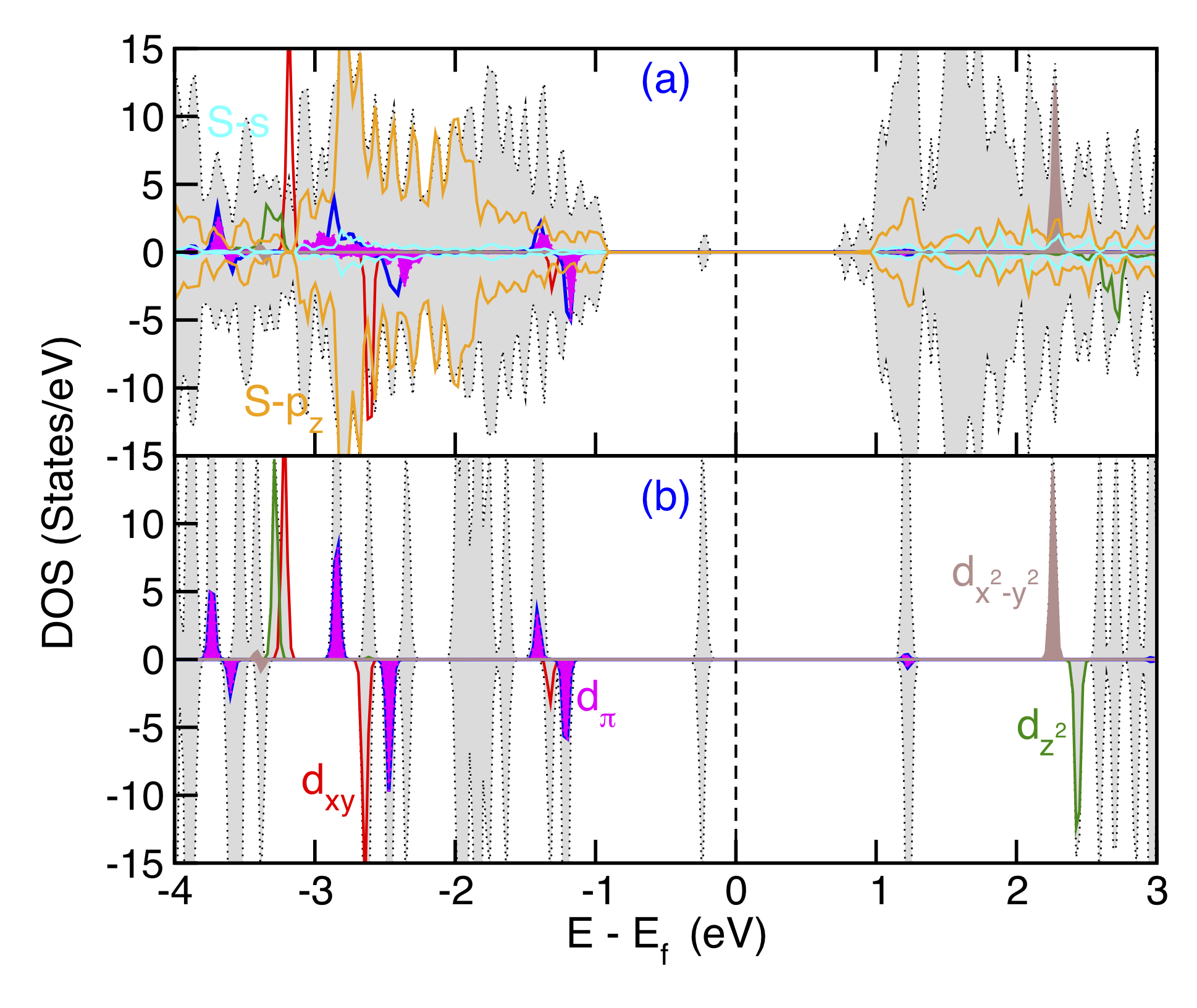}
\caption{(Color online) (a) Total density of states (grey shaded region) of CoPc physisorbed  on MoS$_2$, $m_l$ projected Co-$d$ states of CoPc and $s$, $p_z$ orbitals of top layer S atoms of \ce{MoS2}. The total density of states is scaled down by a factor of 10. (b) Total density of states (grey shaded region) and $m_l$ projected Fe-$d$ states for CoPc in gas phase. \\~\\ }
\label{fig:copc-dos}
\end{figure}
\begin{figure}[tbp]
\centering
\includegraphics[scale=0.4]{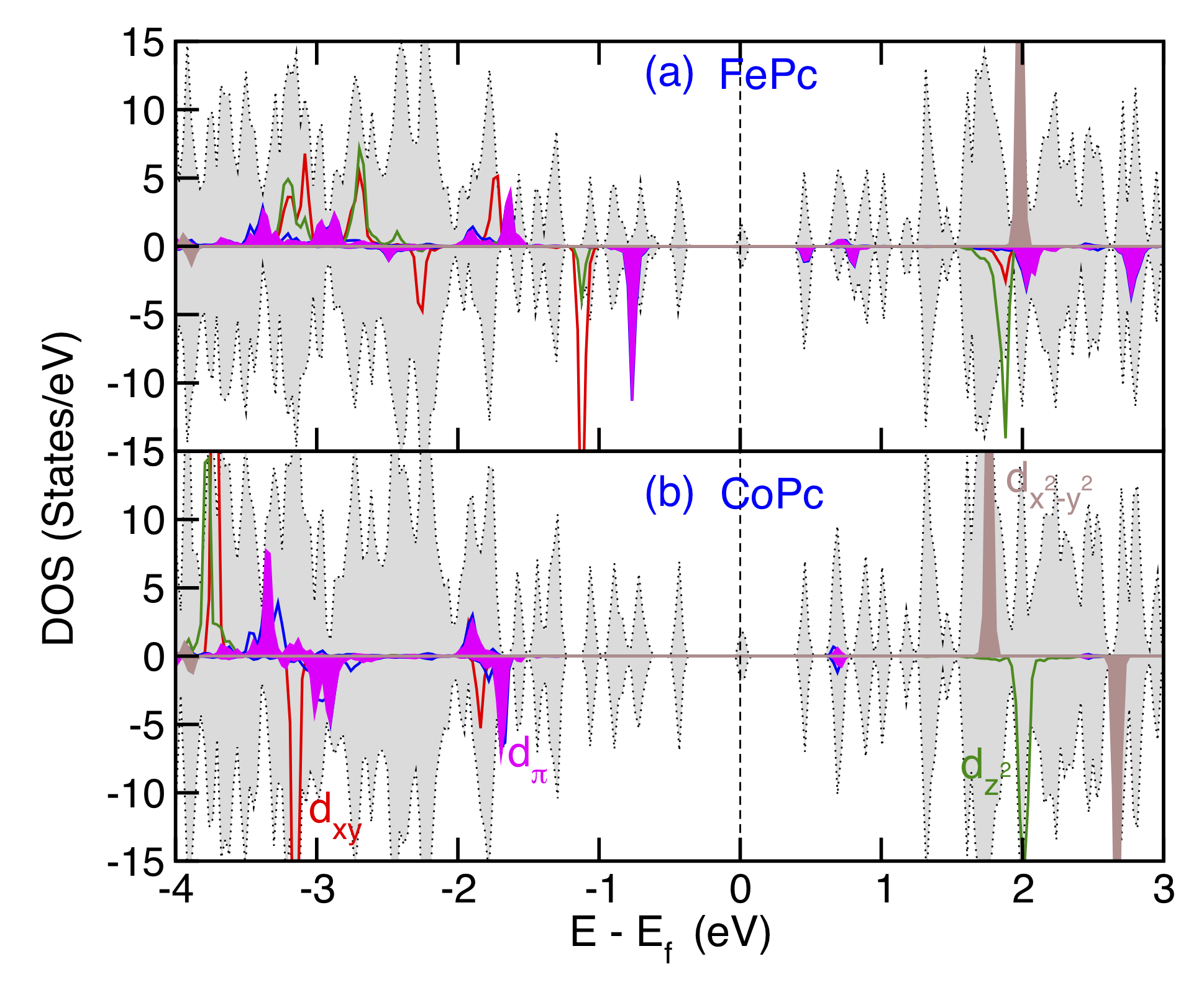}
\caption{(Color online) Total density of states (grey shaded region) of \emph{M}Pc ((a) FePc and (b) CoPc) physisorbed  on Graphene and $m_l$ projected $M-d$ states of \emph{M}Pc. Both total density of states are scaled down by a factor of 10. \\~\\}
\label{fig:on-gr-dos}
\end{figure}
In this subsection, we will discuss the electronic properties of the \emph{M}Pc molecules physisorbed on \ce{MoS2} and graphene. We have used total density of states, site and $m_l$ projected densities of states for our analysis.

In panel (a) of Fig.~\ref{fig:fepc-dos} and Fig.~\ref{fig:copc-dos}, respectively, we have shown the total density of states for FePc and CoPc physisorbed on 2D \ce{MoS2} surface  along with the $m_l$ decomposed $d$ states of the metal center of the \emph{M}Pc molecules. We have also plotted the projected density of states of $p_z$ and $s$ orbitals originating from the top S atom layer of \ce{MoS2}. In order to compare with the gaseous phase of the molecule, we have also plotted the metal center $d$ states along with total density of states of gaseous FePc and CoPc molecule in the panel (b) of the Fig.~\ref{fig:fepc-dos} and Fig.~\ref{fig:copc-dos} respectively. 
In Fig.~\ref{fig:on-gr-dos}(a) and Fig.~\ref{fig:on-gr-dos}(b) respectively, we have plotted the total density of states of FePc and CoPc physisorbed on graphene surface along with the $m_l$ decomposed $d$ states of the metal atom center. 

From the analysis of projected density of states, it can be seen that the out-of-plane $d$ orbitals of the metal atoms become broaden after physisorption on \ce{MoS2}
as compared to the gaseous phase (see Fig.~\ref{fig:on-gr-dos}(a)). These are mainly seen on the occupied out-of-plane $d_{z^2}$ orbital as well as $d_{xz}$ and $d_{yz}$ orbitals. These broadenings of orbitals are mainly due to the hybridization between out-of-plane $d$ orbitals of the physisorbed molecule and the S-$z$ orbitals of top S atoms of \ce{MoS2}. In case of graphene, a similar broadening can also be seen (see Fig.~\ref{fig:on-gr-dos}(b)). However, the amount of hybridization is relatively weak which is evident from the smaller physisorption energy and larger vertical distance (see table~\ref{tab:energy_mag_gr}). The total magnetic moments after physisorption on \ce{MoS2} and graphene are 2.0 $\mu_B$ and 1.0 $\mu_B$ respectively for FePc and CoPc. The majority of magnetic moments are coming from the $d_{z^2}$ orbital of the metal atom. The physisorbed FePc has a spin configuration of S=1. S=1 is equivalent to a 231 electronic configurations where $b_{2g} (d_{xy})$ orbital has occupancy 2, $e_g (d_{xz}, d_{yz})$ orbital has occupancy 3 and $a_{1g} (d_{z^2})$ orbital has occupancy 1. However, for CoPc the spin configuration is S=1/2, which is equivalent to a 241 electronic configurations where $b_{2g}$ orbital has occupancy 2, $e_g$ orbital has occupancy 4 and $a_{1g}$ orbital has occupancy 1. The magnetic moments and the electronic configurations of \emph{M}Pc molecules in the physisorbed situation do not change from the gas phase value significantly. This is because  the hybridized $d$ orbitals are quite below from the Fermi energy. Thus there are no creation of bonding and anti-bonding orbital and hence no weight transfer. Therefore, the electronic configuration and magnetic moments remain same. 


\subsubsection{\label{subsec:spindipole}Spin dipole contribution} 
As mentioned in the introduction, for a low symmetry structure, deformation in the spin densities are expected to be large. These deformed spin densities lead to a large value of the spin dipole moment 7$\langle  T_z \rangle$ where 7$\langle  T_z \rangle$ is the expectation value of the $z$ component of the spin-dipole operator $T$.~\cite{sumantaprl} The discussions of the spin dipole moments are also quite relevant for XMCD measurements, where the measured effective moments contain both spin and spin-dipole moments.

\begin{table}[hbtp]
	\centering
	\caption{Total spin magnetic moment 2$\langle  S_z \rangle$ in {\uB}, spin dipole moment 7$\langle$T$_z$$\rangle$ in {\uB} and effective spin moment $S_{eff}$ = 2$\langle  S_z \rangle$ + 7$\langle  T_z \rangle$ in {\uB}  for $M$Pc molecules physisorbed on monolayer \ce{MoS2} (top table), graphene (middle table) and gas phase $M$Pc (bottom table).}
	\label{table:tz}
	\begin{tabular}{c c c c}
				\rowcolor{Gray}
		System &  2$\langle  S_z \rangle$   &  7$\langle$T$_z$$\rangle$ & $S_{eff}$ \\ \hline
		FePc@\ce{MoS2} & 2.0 & -2.09 & -0.09 \\
		CoPc@\ce{MoS2} & 1.0 & -1.69 & -0.69 \\ 
		\hline
	\end{tabular}\\
	\begin{tabular}{c c c c}
		FePc@graphene & 2.0 & -2.05 & -0.05 \\
		CoPc@graphene & 1.0 & -1.70 & -0.70 \\ 
		\hline
	\end{tabular}\\
	\begin{tabular}{c c c c}
		FePc & 2.0 & -1.99 & 0.01 \\
		CoPc & 1.0 & -1.70 & -0.70 \\ 
		\hline
	\end{tabular}%
\end{table}

Following the formalism of Oguchi \etal ~\cite{oguchi}, the spin dipole operator can be defined as: 
\begin{equation}
T=\sum_{i}Q^{(i)}s^{(i)},
\end{equation}
where, $Q^{(i)}$ is the quadrupolar tensor and can be described as:   
\begin{eqnarray}
Q^{(i)}_{\alpha \beta}=\delta_{\alpha \beta}-3\hat{r}^{(i)}_{\alpha}\hat{r}^{(i)}_{\beta}
\end{eqnarray}    
Every component of $T$ can be written in second quantization form as:   
\begin{eqnarray}
T_{\pm}=T_x\pm iT_y=\sum_{\gamma \gamma^{'}}T^{\pm}_{\gamma \gamma^{'}}a^{\dagger}_{\gamma}a_{\gamma^{'}}
T_z=\sum_{\gamma \gamma^{'}}T^{z}_{\gamma \gamma^{'}}a^{\dagger}_{\gamma}a_{\gamma^{'}}
\end{eqnarray}  
The matrix elements of $T_{\pm}$ and $T_z$ are :
\begin{eqnarray}
T^{\pm}_{\gamma \gamma^{'}}=\bra{\gamma}c^{2}_0s_{\pm}-\sqrt{6}c^{2}_{\pm2}s_{\mp}\pm \sqrt{6}c^{2}_{\pm1}s_{z}\ket{\gamma^{'}}\\
T^{z}_{\gamma \gamma^{'}}=\bra{\gamma}-\sqrt{\frac {3}{2}} c^{2}_{-1}s_{+}+\sqrt{\frac {3}{2}}c^{2}_{1}s_{-}-2c^{2}_0s_{z}\ket{\gamma^{'}}
\end{eqnarray}
where, $\ket{\gamma}=\ket{lm,\sigma}$. 

To calculate the spin dipole moments, we have followed the method prescribed by Freeman \etal~[Ref. \onlinecite{freeman}] and performed density functional theory calculations including all the above discussed effects. The results of calculated spin dipole moments for \emph{M}Pc gas phase molecules and \emph{M}Pc physisorbed on \ce{MoS2} and graphene are tabulated in Table~\ref{table:tz}. As seen from the table, the calculated values of spin dipole moments are -2.09  $\mu_B$ and -1.69 $\mu_B$ for FePc and CoPc physisorbed on  monolayer \ce{MoS2}, respectively.  For graphene, the values are -2.05 $\mu_B$ and -1.70 $\mu_B$ respectively. From our calculation, it is clear that the values of spin dipole moments (7$\langle  T_z \rangle$) are very strong and they are opposite to the spin moment (2$\langle  S_z \rangle$). Hence, the effective spin moment defined as $S_{eff}$ = 2$\langle  S_z \rangle$ + 7$\langle  T_z \rangle$, will be reduced by a huge amount.

Our calculation also indicates that the comparative values of spin dipole moments do not change significantly for both \emph{M}Pc molecules after physisorption on \ce{MoS2} or graphene as compared to the gas phase. The values of spin dipole moments depend mainly on the projected occupancies of the $d$ orbitals of the metal atom center of \emph{M}Pc molecules. From the detailed analysis of density matrix it can be seen that the occupations of these $d$ orbitals do not change significantly in physisorption situation as compared from the gas phase. Hence the changes in spin dipole moments are minimal. 

\subsubsection{\label{subsec:mae}Orbital moments and magnetic anisotropy}

As discussed the importance of magneto crystalline anisotropy in the introduction, we have also incorporated the spin-orbit coupling in the Hamiltonian of our DFT calculations to calculate orbital moments and magneto crystalline energy. 
The magnetic crystalline anisotropy originates from the coupling between lattice and spin. It can be expressed as follows: 
\begin{equation}
H_{SO} =  \zeta(r)\textrm{\textbf{L $\cdot$ S}} = \zeta(r)(L_xS_x+L_yS_y+L_zS_z)
\end{equation}
where $H_{SO}$ is the spin-orbit coupling Hamiltonian and $\zeta$ is the spin orbit coupling constant which can be defined as follows
\begin{equation}
\zeta = \langle \zeta(r) \rangle = \int R^{2}_{3d}(r) \zeta(r)r^2 dr
\end{equation}
where $R_{3d}(r)$ is the radial part of the 3d wave function. The value of $\zeta$ for Fe$^{2+}$ and Co$^{2+}$ have been taken from literature and considered as 49.6 meV and 63.8 meV respectively. \cite{soccref} Using second order perturbation theory, one can write the spin-orbit contribution to the energy as follows
\begin{equation}
\label{eq:eso}
E_{SO} = -\zeta^2 \sum_{u,o}\frac{[\bra{u}\textrm{\textbf{L $\cdot$ S}}\ket{o}\;\bra{o}\textrm{\textbf{L $\cdot$ S}}\ket{u}]}{E_u - E_o}
\end{equation}
In the above equation, $\ket{o}$ and $\ket{u}$ denotes the weighted occupied and unoccupied states of Fe/Co $d$ orbitals. These occupied and unoccupied states are weighted by the occupation of respective $d$ orbitals, $\ket{o/u} = \ket{lm,\sigma}$. The orbital and spin operators are denoted by \textbf{L} and \textbf{S}. $E_o$ and $E_u$ denote the eigenvalues of occupied and unoccupied stated and this values have been taken from the \emph{ab initio} calculations. From the equation~\ref{eq:eso}, it is clear that the spin-orbit coupling energy contribution increases when $E_u - E_o$ decreases. Therefore the relative placement of the $d$ orbitals of the Fe/Co will dictate the energy contribution to the spin-orbit coupling. The relative arrangement of the $d$ orbitals can be affected by the physisorption site of the \emph{M}Pc on \ce{MoS2} or graphene. 

\begin{table}[hbtp]
	\centering
	\caption{MAE-Data within variational approach as implemented in \textsc{vasp} code, magnetic anisotropy energy ($\Delta E = E^{hard}-E^{easy}$) and  orbital moment ($\mu_{orb}$) for $M$Pc molecules physisorbed on monolayer MoS$_2$ (top table),  on graphene (middle table) and gas phase $M$Pc (bottom table).}
	\label{table:mae_var}
	\begin{tabular}{c c c c}
				\rowcolor{Gray}
		&     & $\Delta E$ & Easy \\
				\rowcolor{Gray}
		System & $\mu_{orb}$ &  ($\mu$eV)       & axis \\ \hline
		FePc@\ce{MoS2} & 0.012 & 52 & Out-of-plane \\
		CoPc@\ce{MoS2} & 0.036 & 75 & In-plane  \\
		\hline
	\end{tabular}\\
	\begin{tabular}{c c c c}
				\rowcolor{Gray}
		&     & $\Delta E$ & Easy \\
				\rowcolor{Gray}
		System & $\mu_{orb}$ &  ($\mu$eV)       & axis  \\ \hline
		FePc@graphene & 0.012 & 48.5 & Out-of-plane \\
		CoPc@graphene & 0.037 & 76 & In-plane \\
		\hline
	\end{tabular}\\
	\begin{tabular}{c c c c}
				\rowcolor{Gray}
		&     & $\Delta E$ & Easy \\
				\rowcolor{Gray}
		System & $\mu_{orb}$ &  ($\mu$eV)       & axis \\ \hline
		FePc  & 0.012 & 30 & Out-of-plane \\
		CoPc & 0.037 & 78 & In-plane \\
		\hline
	\end{tabular}%
\end{table}
\begin{table}[hbtp]
	\centering
	\caption{MAE-Data within 2$^{nd}$ order perturbation approach, magnetic anisotropy energy ($\Delta E = E^{hard}-E^{easy}$) and orbital moment ($\mu_{orb}$) for $M$Pc molecules physisorbed on monolayer MoS$_2$ (top table),  on graphene (middle table) and gas phase $M$Pc (bottom table).}
	\label{table:mae_per}
	\begin{tabular}{c c c c}
				\rowcolor{Gray}
		&     & $\Delta E$ & Easy\\
				\rowcolor{Gray}
		System & $\mu_{orb}$ &  ($\mu$eV)       & axis \\ \hline
		FePc@\ce{MoS2} & 0.019 & 103 & Out-of-plane \\
		CoPc@\ce{MoS2} & 0.028 & 58 & In-plane \\ 
		\hline
	\end{tabular}\\
	\begin{tabular}{c c c c}
				\rowcolor{Gray}
		&     & $\Delta E$ & Easy \\
				\rowcolor{Gray}
		System & $\mu_{orb}$ &  ($\mu$eV)       & axis \\ \hline
		FePc@graphene & 0.024 & 107.8 & Out-of-plane \\
		CoPc@graphene & 0.030 & 60.2 & In-plane \\ 
		\hline
	\end{tabular}\\
	\begin{tabular}{c c c c}
				\rowcolor{Gray}
		&     & $\Delta E$ & Easy \\
				\rowcolor{Gray}
		System & $\mu_{orb}$ &  ($\mu$eV)       & axis \\ \hline
		FePc & 0.026 & 98 & Out-of-plane \\
		CoPc & 0.028 & 60 & In-plane \\ 
		\hline
	\end{tabular}%
\end{table}

We have used two different approach to calculate the magneto crystalline energy -- i) variational approach implemented in \textsc{vasp} code and ii) 2$^{nd}$ order perturbation approach. We have tabulated the result of these two approaches in table~\ref{table:mae_var} and in table~\ref{table:mae_per} respectively. We have reported the orbital magnetic moments, magnetic anisotropy energies along with the information about the easy axis in table~\ref{table:mae_var} and~\ref{table:mae_per}. Both approaches mentioned before gives similar qualitative results. As seen from the table~\ref{table:mae_var}, the orbital moments values do not change significantly in the physisorbed systems. The analysis of $m_l$ projected density of states (see Fig.~\ref{fig:fepc-dos}, Fig.~\ref{fig:copc-dos}, Fig.~\ref{fig:on-gr-dos}) indicate that although the $d$ orbitals below the Fermi levels are broadened in the physisorbed system compared to the gas phase, the band center remains same . Hence the orbital moments remain almost similar. The relative positions of $d$ orbitals of Co metal center in CoPc do not change significantly from gas phase in to the physisorbed phase. Therefore, the transition matrices are similar and hence the magneto crystalline anisotropy energies are similar in values. However, for FePc the values of magneto crystalline anisotropy energies are different from the gas phase as the Fe-$d$ orbitals overlapped differently in the unoccupied regions in physisorbed cases. 
Our results indicate that  
while for CoPc the easy axis of magnetization is in-plane, the easy axis is out-of-plane for FePc.    

\subsubsection{\label{subsec:workfunc}Work function}
Decoration by the molecules is one of the method to change the surface work function. These changes can enable the use of graphene or \ce{MoS2} in designing various different nano-devices.   In order to investigate possible change in work function ($\Phi$) of the \ce{MoS2} and graphene due to the \emph{M}Pc physisorption, we have calculated the work function of the pristine monolayer \ce{MoS2}, graphene and \emph{M}Pc physisorbed \ce{MoS2} and graphene. The work function $\Phi$ is defined as the minimum energy required to remove an electron from a material to the vacuum. Hence the work function $\Phi$ can be defined as the following,
\begin{equation}
\label{eq:workfunction}
\Phi = V(\infty) - E_f
\end{equation}
where, $V(\infty)$ is the self-consistent electrostatic potential in the vacuum far from the surface and $E_f$ is the Fermi energy and is calculated from a ground state self-consistent calculation. Here we have taken the Fermi energy to be the top of valence band. The surface work functions were calculated by applying the Neugebauer-Scheffler dipole correction~\cite{workfunction_prb} for the direction perpendicular to the surface. The electrostatic potential is obtained as function of $z$ by averaging the self-consistent potential parallel to the surface, i.e., in the $xy$ plane. Then the potential $V(\infty)$ is approximated as the value of potential at the vacuum layer where $V(z)$ is reaching its asymptotic limit.

\begin{table}[hbtp]
	\centering
	\caption{Comparison of work function ($\Phi$) values for monolayer \ce{MoS2}, Graphene and $M$Pc molecules physisorbed on \ce{MoS2} and Graphene.\\~\\}
	\label{table:workfrunc}
	\begin{tabular}{c c c}
		\multicolumn{2}{c}{System} & $\Phi$ (eV) \\ \hline
		\multicolumn{2}{c}{\ce{MoS2} (1L)} & 5.94 \\ \hline
		\multirow{2}{*}{\ce{MoS2}} & FePc & 4.94 \\
		& CoPc & 5.17 \\ \hline
	\end{tabular}
	\begin{tabular}{c c c}
		\multicolumn{2}{c}{System} & $\Phi$ (eV) \\ \hline
		\multicolumn{2}{c}{Graphene (1L)} & 4.23 \\ \hline
		\multirow{2}{*}{Graphene} & FePc & 4.25 \\
		& CoPc & 4.25 \\ \hline
	\end{tabular}%
\end{table}
In table~\ref{table:workfrunc}, we have tabulated the calculated values of $\Phi$ for the above mentioned cases. The value of $\Phi$ for monolayer graphene is 4.23 eV. This value is quite consistent with previous published result.~\cite{grwf} For \ce{MoS2}, the computed value of $\Phi$ is 5.94 eV which is similar to earlier published result.~\cite{mos2wf1} It is evident from our calculation is that the value of $\Phi$ does not change significantly after \emph{M}Pc physisorption on graphene. However, the work function value of \ce{MoS2} decreases after \emph{M}Pc physisorption. Hence our calculation shows that it can be possible to tune the $\Phi$ of \ce{MoS2} by \emph{M}Pc physisorption. Work function dependency on \emph{M}Pc physisorption on \ce{MoS2} opens up the possibility of using these materials in electronic and optoelectronic devices. 

\section{\label{sec:conclusion}Conclusions}
We have performed detailed investigation of electronic and magnetic properties of \emph{M}Pc (M = Fe, Co) physisorption on \ce{MoS2} and graphene using density functional theory along with Coulomb correlation for the metal $d$ electrons. We have considered various possible physisorption sites both on \ce{MoS2} and on graphene to find out energetically most favorable configuration. From our calculation, we have found out that \emph{M}Pc molecules are physisorbed strongly on \ce{MoS2} as compared to  graphene. The most stable physisorption site on \ce{MoS2} is when the metal atom center of the \emph{M}Pc molecules resides on top of S atom. The other physisorption sites have much higher energy ($\sim$ 0.14 eV). However, on graphene, the physisorption energies are very similar for different physisorption sites. The out-of-plane $d$ orbitals of the metal centers hybridize with the out-of-plane orbitals from \ce{MoS2} and graphene. The magnetic moment comes from $d_{z^2}$ orbital of the metal atom. \emph{M}Pc physisorption reduces the work function of \ce{MoS2} by $\sim$ 1 eV whereas it does not change much in the case of graphene. \emph{M}Pc physisorption causes a large spin dipole moment opposite to the spin moment, which can be measured by XMCD experiment. It causes a huge reduction of effective magnetic moments of the system. Our calculations of magnetic anisotropy energies using both variational approach and $2^{nd}$ order perturbation approach indicate no significant changes in the magnetic anisotropy energy values after physisorption of the \emph{M}Pc molecules. An out-of-plane magnetic anisotropy can be observed in the case of FePc whereas for CoPc it is in-plane. 

\section*{Acknowledgment} 
SH and BS would like to acknowledge KAW foundation for financial support. In addition, BS acknowledges Carl Tryggers Stiftelse, Swedish Research Council and KOF initiative of Uppsala University for financial support. We are grateful to NSC under Swedish National Infrastructure for Computing (SNIC) and the PRACE-2IP project resource Cy-Tera supercomputer based in Cyprus at the Computation-based Science and Technology Research Center (CaSToRC) and Salomon cluster based in Czech Republic at the IT4Innovations for computer hours. Structural figures are generated using VMD.~\cite{vmd} 

\bibliography{biblio,references}

\end{document}